\documentstyle[12pt]{article}

\newcommand{\NP}[1]{Nucl. \ Phys.}
\newcommand{\PL}[1]{Phys. \ Lett.}
\newcommand{\PRL}[1]{Phys.\ Rev.\ Lett. }
\newcommand{\AP}[1]{Ann.\ Phys. }
\newcommand{\IM}[1]{Inv.\ Math. }
\newcommand{\JMP}[1]{ J.\ Math.\ Phys. }
\newcommand{\RMP}[1]{Rev.\ Mod.\ Phys.}
\newcommand{\JDG}[1]{ J.\ Diff.\ Geom. }
\newcommand{\PTP}[1]{ Prog.\ Theor.\ Phys. }
\newcommand{\SPTP}[1]{Suppl.\ Prog.\ Theor.\ Phys. }
\newcommand{\PR}[1]{Phys.\ Rev. }
\newcommand{\PREP }[1]{Phys.\ Reports }
\newcommand{\NC}[1]{Nuovo \ Cim. }
\newcommand{\NCL }[1]{Nuovo\ Cim.\ Lett. }
\newcommand{\CMP}[1]{Commun.\ Math.\ Phys. }
\newcommand{\TMF}[1]{Theor.\ Math.\ Phys. }
\newcommand{\CQG}[1]{Class.\ Quant.\ Grav. }
\newcommand{\FAA}[1]{Funct. Analys. Appl. }
\newcommand{\JP}[1]{J.\ Phys. }
\newcommand{\JA }[1]{ J. Algebra }
\newcommand{\JFA}[1] {J. Funct. Anal. }
\newcommand{\MPL}[1] { Mod. Phys. Lett. }
\newcommand{\IJMP}[1] { Int. J. Mod. Phys. }
\newcommand{\RMAP}[1] { Rep.\ Math.\ Phys. }

\begin{document}

\centerline{\large \bf  Dual  Supergravity in D=10, N=1 Superspace}

\bigskip
\centerline{\large \bf with Tree-Level Superstring Corrections}

\vspace*{1cm}

\centerline{\large M.V.Terentjev\footnote{E-mail address:
TERENT@vxitep.itep.msk.su}}

\bigskip

\centerline{Institute of Theoretical and Experimental Physics (ITEP)}
\centerline{\it Moscow, 117259, Russia}

\bigskip

\begin{abstract}
The dual version of the D=10 N=1 supergravity (SUGRA) is
considered in the superspace approach. The  superstring (anomaly
compensating) corrections are described by the 3-form superfield
$A_{abc}$ .  The complete set of dynamical equations  for the
$A$-field and for physical fields of the theory are presented.
The solution of the $A$-field equations as a finite order
polynomial in terms of curvature and graviphoton superfields is
given. It makes possible to incorporate  some of the superstring
corrections in the dual supergravity in the explicit and closed
form.
\end{abstract}

\bigskip

{\bf 1. \, INTRODUCTION}

\bigskip
It was shown by Gates and Nishino \cite{GN1} that the "dual"
version of the D=10, N=1 supergravity (IB-type theory) is an
alternative to the "usual"  version (type IA)  not only as the
field theory, but also as a classical background of the D=10,
N=1 superstring theory. First, these two versions  (IA and IB)
were considered by Chamseddine \cite{C}.  Interaction with
matter was constructed in \cite{BRWN}, \cite{CM} (for  IA
theory) and in \cite{GN2} (for IB theory).

However, both the  IA and IB versions, considered as an
effective theory  of massless modes of the type I (heterotic)
superstring, can not be described by the lagrangian of  ref's
\cite{C}, \cite{BRWN}-\cite{GN2}. That is because these
supergravity versions are anomalous, while the heterotic
string theory is not.

The Green-Schwarz \cite{GS} mechanizm was invented for
cancellation of the supergravity anomalies. It was implemented
originally to the IA-theory, but it works also in the IB theory
(cf. \cite{GN2}).  According to \cite{GS}, the Lorentz-type
Chern-Symons terms are needed to compensate anomalies. Such
terms are added to the graviphoton field-strength in the IA
theory, and (in some approximations) immediately to the
lagrangian as special interaction terms in the IB theory. After
this modification the IA and IB supergravities are
no longer supersymmetric.
Restoration of supersymmetry is an important long standing
problem.

The  implicit solution of the supersymmetrization problem for
some of the anomaly compensating terms in the IA theory was
presented in the refs. \cite{AFRR},\cite{BBLPT}.  (See
ref.\cite{MANY}, where the present state of the problem is
described and more detailed list of references is given). The
explicit solution may be  obtained from the results of
\cite{AFRR}, \cite{BBLPT} as an infinite series in the string
slope parameter. In general, this series contains all positive
integer powers of the graviphoton field-strength and the
curvature tensor.

We demonstrate in this paper, that the situation is different
and more simple in the IB ("dual") version of supergravity. The
supersymmetrization may be achieved by adding  a finite number
of terms, which are powers of the graviphoton (7-form)
field-strength and the curvature tensor.  \footnote{The IA and
IB supergravities are completely equivalent in the zero order in
the string coupling constant, i.e. in the case, where
Green-Schwarz mechanizm is "switched off". In the general case,
when powers $ H^n, \ \ n\geq 3$ of the graviphoton field are
included, these theories may be different. The possible
connection between them is discussed in \cite{BR1}, where one
may find other references on the subject.} This result provides
the  explicit solution of supersymmetrization problem in the
dual supergravity, obtained in the same approximations as in the
ref's  \cite{AFRR},\cite{BBLPT}.   Some preliminary results were
obtained in the refs.\cite{GN3} - \cite{T2}. Our derivation here
is based on the remarks made by D'Auria and Fre
\cite{AF}. We use also general results, obtained in
\cite{BBLPT}.

We consider the case of pure supergravity in the D=10, N=1
superspace. (The incorporation of matter degrees of freedom is
the standard procedure).  This  work is based on the previous
paper \cite{T3}, where   dynamical equations for superstring
corrections were described using the simple parametrization of
Bianchi Identities (BI's), introduced by Nishino \cite{N}.  Our
notations correspond in general to that of ref. \cite{T3} (they
are described in \cite{BBLPT}, \cite{T1}, \cite{T2}), but there
are small differences which are self-evident.

       \bigskip

{\bf 2. \, SOLUTION OF TORSION BIANCHI IDENTITIES}

\bigskip

First, we show, that all the supergravity equations of motion
(e.m.) in superspace follow from the supertorsion BI's. The only
difference between usual (type IA) and "dual" (IB) versions is
in the connection between components of supertorsion and
graviphoton superfields.

The standard set of torsion BI's is:

$$ D_{[A}{T_{BC)}}^D + {T_{[AB}}^Q\, {T_{QC)}}^D  +
{R_{[ABC)}}^D  = 0.                 \eqno(2.1) $$

Here the summation $Q$-index is not included in the
symmetrization procedure.  Covariant derivatives $D_A$ obey the
commutation (anticommutation) relations:

 $$ (D_A\, D_B - (-1)^{ab}D_B\, D_A) = - {T_{AB}}^Q\, D_Q +
 R_{ABcd}\, M^{cd}    \eqno(2.2)  $$

 where $T_{AB}^C $ is the torsion superfield,  ${R_{ABC}}^D  $
is the supercurvature,  $M^{cd}$ is the generator of Lorentz
transformations: ${(M^{cd})_\alpha}^\beta = {1\over
4}{(\Gamma^{cd})_\alpha}^\beta$ in the spinorial representation,
$ {R_{AB\alpha}}^ \beta = {1\over 4}
R_{ABcd}{(\Gamma^{cd})_\alpha}^\beta. $

We use the constraints \cite{N}:

$$ {T_{\alpha\beta}}^c = \Gamma_{\alpha\beta}^c \, , \ \ \
 T_{\alpha\beta}^\gamma = T_{\beta b}^c = 0, \ \ \
T_{a\beta}^\gamma ={({\hat X}\Gamma_a)_\beta}^\gamma \, ,
                                                           \eqno(2.3) $$

where ${\hat X} \equiv X_{abc}\Gamma^{abc}$.  This
parametrization provides the complete solution of the BI's (2.1)
in terms of $T_{ab}^c$  and  $T_{ab}^\gamma$ -components.

It follows from (2.1):

 $$ X_{abc} = {1\over 72}\, T_{abc}\, , \ \ \ D\, T_{abc} = - {1\over 2}\,
 {\Gamma_{abc}}^{ij}\, T_{ij}\, ,
                                                                   $$
  $$                       D^c\, T_{abc} =0,
                                                             \eqno(2.4) $$

where $T_{abc} \equiv T_{ab}^d \eta_{dc} $ is the completely antisymmetric
tensor. Furthemore:

$$  T_{ab}\Gamma^{ab} =0, \ \ \
 DT_{ab} =  D\Gamma_{[a}L_{b]} = 0,  $$

$$ D\Gamma_{[ab}T_{cd]} =  {16\over 3} D_{[a}T_{bcd]} +
   {20\over 3}T^2_{[abcd]} \, , $$
$$    D\Gamma_{[abc}L_{d]} = -8\, D_{[a}T_{bcd]} -
{40\over 3}T^2_{[abcd]} \, ,                                  \eqno(2.5)  $$

where $L_a \equiv T_{ab}\Gamma^b, \ \ \ L_a\, \Gamma^a =0   $.
We do not write spinorial indices explicitely in the cases when
their position may be reconstructed unambiguosly.  Here and
below we use the notations:

$$TA =
T_{ijk}A^{ijk}, \ \ (TA)_{ab} = T_{aij}{A_b}^{ij}, \ \ (TA)_{abcd} = T_{abj}
{A_{cd}}^j                                                      \eqno(2.6) $$.

where $A_{abc}$ is an arbitrary 3-form superfield, $ T^2 = TT $.

The components of the curvature tensor are:

$$  R_{\alpha\beta ab} = - {5\over6}\, T_{abc}\Gamma^c_{\alpha\beta} -
{1\over 36}\, T_{ijk}\, ({\Gamma^{ijk}}_{ab})_{\alpha\beta} ,
                                                              \eqno(2.7)  $$,
 $$ R_{abc} = - 2\, T_{a[b}\Gamma_{c]}
 + {3\over 2} L_{[a}\Gamma_{bc]} ,
                                                              \eqno(2.8)  $$
  $${ R_{ab}}^{cd} = - {1\over 8}D\Gamma^{cd}T_{ab} - {1\over
 3}D_{[a}{T_{b]}}^{cd} - {1\over 108}\delta^{cd}_{[ab]} T^2 + {1\over
 18}\delta_{[a}^{[c}{(T^2)^{d]}}_{b]} - {1\over 6} {(T^2)_{ab}}^{cd}
                                                               \eqno(2.9)  $$
  In particular:  $$  R_{[abc]d} = - D_{[a}T_{bc]d} -
   T^2_{[abc]d} , \ \ \ R_{[ab]} = 0 , \ \  \    R = - {1\over 3}\, T^2 ,
                                                            \eqno( 2.10) $$
    $$
  R_{ab} = - {1\over 8} D\Gamma_{(a}L_{b)} - {1\over 36}T^2 \eta_{ab} -
 {1\over 18} T^2_{ab} ,
                                                           \eqno(2.11 )$$

  where $R_{ab} \equiv {R_{acb}}^c$  is the Ricci tensor, $R
\equiv R_{ab}\eta^{ab}$ is the curvature scalar,   $
(R_{abc})_\beta \equiv R_{\beta abc} $.

Now we introduce the dilaton $\phi$, and define the dilatino
$\chi_\alpha$-field by:

$$ \chi \equiv D\phi                                       \eqno(2.12) $$

The most general expression for the dilatino field spinorial derivative
may be taken in the form:

 $$ D_\alpha \chi_\beta = -{1\over2}{\hat D}_{\alpha \beta}\phi
 +(-{1\over 36} \phi T_{abc} + A_{abc})\, \Gamma^{abc}_{\alpha \beta},
                                                            \eqno(2.13) $$

where $ \hat D \equiv D_a\Gamma^a $, but $A_{abc}$ is a new superfield.

 The D=10 supergravity multiplet contains $\phi \vert$, $\chi \vert$ -fields
($\phi \vert $ is the first component of $\phi$- superfield, etc.),
gravitino $\psi_m^\gamma \vert$ field (note, that $T_{mn}^\gamma =
D_{[m}\psi_{n]}^\gamma$ ), graviton $E_m^a \vert $ and graviphoton
 3-form field  $H_{abc}\vert $  or 7-form field $N_{a_1 \ldots a_7}\vert  $
in the dual supergravity, which are related to $T_{abc}\vert $ (see below).
All the equations of motion for these fields follow from (2.1), (2.2) and
(2.13) if the constraints (2.3) are imposed and the subsequent
formulas (2.4) - (2.11) are used.

The gravitino e.m. may be derived, calculating the quantity
$(D\Gamma_aD)\chi_\beta$ with the help of (2.2) and (2.13). It takes the
form:

 $$ Q_a \equiv \phi L_a - D_a\chi - {1\over 36} \Gamma_a {\hat
T}\chi - {1\over 24} {\hat T}\Gamma_a\chi + {1\over 42} \Gamma_a
\Gamma^{ijk}DA_{ijk} + {1\over 7} \Gamma^{ijk}\Gamma_a DA_{ijk} = 0,
                                                            \eqno(2.14) $$
where $\hat T \equiv T_{abc}\Gamma^{abc}$.
 The
dilatino e.m. follows immediately from the projection
$\Gamma^a Q_a = 0:$

$$ {\hat D}\chi + {1\over 9}{\hat T}\chi +
 {1\over 3}\Gamma^{ijk}DA_{ijk} = 0.
                                                              \eqno(2.15)  $$

 The e.m. for the dilaton field  follows from the explicit
 calculation of the spinorial derivative  $D\Gamma^aQ_a = 0$.
  The result is:

$$ D_a^2 \phi + {1\over 18}\phi T^2 - 2\, TA -
 {1\over 24} D\Gamma^{ijk}DA_{ijk} = 0.
                                                             \eqno(2.16)  $$

The graviton e.m. follows from (2.11) if one calculates the
corresponding spinorial derivative $D\Gamma_{(a}L_{b)}$ from the
gravitino e.m.  The result is:

 $$ \phi R_{ab} = - L_{(a}\Gamma_{b)}\chi - {1\over 36}\phi\eta_{ab}T^2 +
 D_{(a}D_{b)}\phi -$$
 $$- 2\, T_{(a}A_{b)} + {3\over 28}D{\Gamma^{ij}}_{(a}DA_{b)ij} -
 {5\over 336}\eta_{ab} D\Gamma^{ijk}DA_{ijk}
                                                            \eqno(2.17)$$

The e.m. $D_{[a}T_{bcd]} = \ldots $  and the similar equation
for  $A_{abc}$ -field may be derived if one combains the 210 IR
projections $D\Gamma_{[abc}Q_{d]}=0  $ and
$D{\Gamma_{abcd}}^jQ_j = 0 $. One obtains in this way:

$$D_{[a}(\phi T_{bcd]}) + {3\over 2} T_{[ab}\Gamma_{cd]}\chi + {3\over 2}
\phi T^2_{[abcd]} +$$
 $$+ {1\over 12} (T\epsilon A)_{abcd} + 6\,  (TA)_{[abcd]}
 + {3\over 4}D{\Gamma_{[ab}}^jDA_{cd]j} = 0
                                                           \eqno(2.18) $$
  and:
   $$ D_{[a}\,
A_{bcd]} + 2\, (TA)_{[abcd]} +{1\over 360}(T\epsilon A)_{abcd}- $$ $$ -
 {1\over 16\cdot 60}D{\Gamma_{abcd}}^{ijk}D\, A_{ijk} + {1\over 16}
 D{\Gamma_{[ab}}^iD\, A_{cd]i} = 0,
                                                             \eqno(2.19) $$
where $(T\epsilon A)_{abcd}
 \equiv T_{ijk}{\epsilon_{abcd}}^{ijkmns} A_{mns}$.

One more equation for the $A_{abc}$- field follows from
the 45 IR projection $D\Gamma_{[a}Q_{b]} = 0$ :

$$ {D\Gamma_{[a}}^{ij}D\, A_{b]ij} + 56\, D^jA_{jab}
- {64\over 3} (TA)_{[ab]} =0
                                                              \eqno(2.20) $$
There are several possible consistency checks of the presented equations.
For example, one may derive the graviton e.m. (2.17) considering the
 projection $D\Gamma_{(a} Q_{b)} = 0$, etc.

  Eventually, we may derive one more equation, considering the
   1200 IR projection of the quantitiy $D_\alpha D_\beta \chi_\gamma$.  We
   obtain \cite{T3}:

   $$ (D_\alpha A_{abc})^{(1200)} = 0
                                                            \eqno(2.21) $$
Note, that (2.19) follows from the eq.(2.21) \cite{T3}.

  In the case $A_{abc} = 0$ the e.m.'s (2.14) - (2.18) and the
last equation in (2.4) correspond (after some field
redefinitions) to the supergravity of ref.\cite{C}. The
incorporation of matter degrees of freedom leads to $A_{abc} =
\lambda \Gamma_{abc} \lambda $, where $\lambda $ is the gluino
field. In this case (2.14) - (2.18), and (2.4) correspond to the
e.m.'s of gravity sector of the refs. \cite{CM} or \cite{GN2}
supergravity.

  Now we return to the more general case of anomaly free
supergravity. The $A_{abc}$-tensor may be found in this case
from the parallel consideration of $N_{a_1 \ldots a_7}  $ and
$H_{abc} $ BI's.

\bigskip

{\bf 3. \, SOLUTION OF GRAVIPHOTON BIANCHI IDENTITIES}

\bigskip

{\bf \it  Dual Supergravity}

\bigskip

The BI's for the 7-form graviphoton field of dual supergravity has the form:

$$ D_{[A_1}N_{A_2 \ldots A_8)} + {7\over2}\, {T_{[A_1A_2}}^Q \,
N_{QA_3\ldots A_8)} \equiv 0                                  \eqno(3.1) $$

(The summation $Q$-index is not included in the symmetrization).
There is no need to introduce anomaly compensating Chern-Symons terms
in (3.1) because the $N$-field appears to be the Lorentz-covariant in
the Green-Schwarz mechanizm.
But there is a freedom in the solution of (3.1), that incorporates the
anomaly compensating contributions in the dual supergravity.

The following set of constraints provides the solution of BI's (3.1):

$$ N_{\alpha\beta a_1 \ldots a_5} =
 - (\Gamma_{a_1\ldots a_5})_{\alpha\beta} ,
                                                           \eqno(3.2) $$
     $$  N_{abc} = T_{abc} ,
                                                            \eqno(3.3) $$
 where
$$ N_{abc} \equiv {1\over 7!} {\epsilon_{abc}}^{b_1\ldots b_7}
\, N_{b_1\ldots b_7}
                                                               \eqno(3.4) $$
All other components of the $N$-field are equal to zero.

This solution is completely consistent with  torsion BI's
in the sec.2. It is important, that no additional restrictions  on the
choice of the $A_{abc}$ - field follow from (3.1). It is still arbitrary
except for the constraints (2.19) - (2.21). Due to the condition (3.3), all
the e.m.'s obtained in the sec.2  become  the equations of dual (type IB)
supergravity. The explicit form of the $A$-field that includes all (anomaly
compensating) superstring corrections may be found from  BI's
of the usual (type IA) supergravity.

\bigskip

{\bf \it Usual Supergravity}

\bigskip

The graviphoton BI's of usual anomaly free supergravity has the form:

$$ D_{[A}H_{BCD)} + {3\over 2}\, T_{[AB}^Q\, H_{QCD)}
- {3\over 2}\, \gamma \, {R_{[AB}}^{ef}{R_{CD)}}_{ef} = 0.
                                                             \eqno(3.5) $$

(The summation $Q$-index is not included in the symmetrization).
The term, proportional to $\gamma$ in (3.5) stems
from the Lorentz Chern-Symons form $\omega_L $ in the definition
of Lorentz-covariant 3-form graviphoton field $H$:
 $H = dB - \gamma \omega_L $, where $B$ is the 2-form graviphoton
 potential.

Our analysis of BI's (3.5) follows closely to the approach of \cite{RRZ},
but it is more simple in the parametrization of sec.2.  The self-consistent
solution of (3.1) has the form:

 $$ H_{\alpha \beta \gamma} = 0,
                                                              \eqno(3.6) $$
 $$ H_{\alpha \beta c} = \phi \, (\Gamma_c)_{\alpha \beta}
    + \gamma \, W_{\alpha \beta c},
                                                               \eqno(3.7) $$
$$ H_{\alpha bc} = - (\Gamma_{bc} \chi)_\alpha +
   \gamma \, W_{\alpha bc},
                                                               \eqno(3.8) $$
$$ H_{abc} = - \phi \, T_{abc} + \gamma W_{abc}
                                                               \eqno(3.9) $$
Furthemore, the equation for the $A_{abc}$ - field follows from (3.5):
 $$ A_{abc} = \gamma L_{abc}
                                                              \eqno(3.10) $$

The functions $W$ and $L$ appear due to the contribution of the
$RR$ term in the BI (3.5).  We present the necessary explicit
expressions for them.

 The $W_{\alpha \beta c}$  is found
from the (0,4) sector of (3.5).  (The (p,q) sector is defined as the
equation with p bosonic and q fermionic external indices). We get:

 $$ W_a = X_{ab}\Gamma^b + X_{a,\,b_1\ldots b_5}\Gamma^{b_1\ldots b_5}
                                                            \eqno(3.11) $$
  where
   $$ X_{ab} = {7\over 9}\, (T^2)_{ab}
                                                           \eqno(3.12) $$
  $$
X_{a,\,b_1\cdots b_5} = {1\over 2\cdot 5!}
 (\delta_{\,b_1\ldots b_5}^{[c_1\ldots c_5]} +
 {1\over 5!}\,{\epsilon_{b_1\ldots b_5}}^{c_1\ldots c_5} )
  \, (G_{a,c_1\ldots c_5} + \Theta_{ac_1\ldots c_5})
                                                               \eqno(3.13) $$
The $G$-tensor is the 1050 IR of O(1.9) (${G^a}_{,ac_2\ldots c_5} = 0$):

$$ G_{a,c_1\ldots c_5} = {40\over 9}(T_{a[c_1c_2}T_{c_3c_4c_5]} -
{1\over 2} {\eta}_{a[c_1} T^2_{c_2\ldots c_5]})
                                                           \eqno(3.14) $$

The $\Theta $ is a totally antisymmetric tensor ( 210 IR of O(1.9)).
It is not fixed at the (0,4) level of BI.

At the (1,3) level we get the  $ W_{\beta ab} $ in the form:

$$ W_{ab} = -
 (\delta_{[ab]}^{\, cd} +
 {1\over 12} \eta^{cd}\Gamma_{ab} +
  {1\over 6}{\Gamma_{[a}}^{(c}\delta_{b]}^{d)})
 (DX_{cd} +
 DX_{c,de_1\ldots e_4}\Gamma^{e_1\ldots e_4} +$$
 $$+ {5\over
 3}R_{cij}{T_d}^{ij} + {1\over 18} R_{c[dm}T_{ijk]}\Gamma^{mijk} )
                                                            \eqno(3.15) $$
At the same (1,3) level we get the restriction on the $D\Theta_{abcd}$ in the
form of the projection to the 1440 IR of O(1.9):
  $$( D\Theta_{abcd} - {4\over 3} L_{[a}T_{bcd]})^{(1440)} = 0
                                                            \eqno(3.16) $$

where

$$ \Theta_{abcd} \equiv {1\over 6!} {\epsilon_{abcd}}^{a_1\ldots a_6}
\Theta_{a_1\ldots a_6}
                                                             \eqno(3.17) $$

 One may find the general solution of this equation. The result is:
 $$ \Theta_{abcd} = - {4\over 9}(D_{[a}T_{bcd]} + 2\, T^2_{[abcd]})
                                                            \eqno(3.18) $$

Now we come to the (2,2) level.
There are two different projections of (3.1) at this level to the 120 IR of
 O(1.9).  The first one produces the relation (3.9), where $W_{abc}$ is a
 complicated tensor, depending on the torsion components and their
 derivatives. The explicit form of $W_{abc} $ is not interesting for us
 here.

 The second projection leads to the eq.(3.10), where $L_{abc}$ is equal to:

 $$ L_{abc} = -{12\cdot 16}\, D^j\Theta_{jabc} - 96\,
 {T^{ij}}_{[a}\Theta_{bc]ij} + 2\,  D{\Gamma^{mn}}_{abc}W_{mn} + $$
  $$ +
 32\cdot 160\,  T^{ijm}X_{m,ijabc} + 128\, T_{[abc}{X_{n]}}^n + {320\over 3}
 {R^{mn}}_{[mn}T_{abc]}+$$ $$ + 2\, {R_m}^{ij}{\Gamma^{mn}}_{abc}R_{nij} +
{8\over 9}(R\epsilon T)_{abc}.
 \eqno(3.19) $$

 where $(R\epsilon T)_{abc} = R_{mnpq}{\epsilon_{abc}}^{mnpqijk} T_{ijk}. $

  All the spinorial derivatives in (3.19) may be calculated
explicitely using the formulas of sec.2. Equation (3.10)
corresponds to the eq.(4.7) from \cite{BBLPT}.  Equation (3.19)
corresponds to the eq.(18) from  \cite{RRZ}. The validity of the
condition (2.21) is garanteed by the main theorem of the
ref.\cite{BBLPT} and may be checked by direct (complicated)
calculations. The validity of (2.19) follows from (2.21)
\cite{T3}. We were not able now to check the validity of (2.20)
because  very long calculations are required for this check. But
we may expect from general grounds that  this equation is also
valid because it is garanteed by the (supposed) self-consistency
of the superspace description of the usual anomaly free
supergravity .

 \bigskip

 { \bf 4.\, CONCLUSION AND MAIN RESULTS}

 \bigskip

Note, that $A_{abc} = 0 $ due to the eq. (3.10) if $\gamma = 0
$. So, the $A_{abc}$-field is connected in this approach only
with the contribution of anomaly compensating terms. One may
easily extract from (3.19)  some special terms   considered
before (see \cite{GN3}-\cite{T1}),  which are explicitely
connected with compensation of anomalies. (They are related with
generic terms $\sim \gamma R^2 $ and $\sim \gamma N \Omega $ in
the lagrangian, where $ d\Omega = R^2 $ ).  The complete
analysis of (3.13) will be presented in the subsequent
publication.

 It is important, that $A_{abc}$ is defined by the eqs.(3.10),
(3.19) quite independently on the $H_{abc}$ - field, so (3.10)
may be used as an anzatz for the $A$-field in the equations of
sec.2 independently on the subsequent choice (IA or IB) of the
supergravity type.

 If we add (3.10) to the e.m.'s of sec.2  and change everywhere
$T_{abc}$ to $N_{abc}$ according to the eq.(3.3), we obtain the
description of the dual supergravity including (anomaly
compensating) superstring corrections.  So, the equations (2.4),
(2.14)-(2.18) together with  (3.3)  and (3.10), (3.19)  provide
the mass shell explicit partial solution of the problem of
supersimmetrization in the dual supergravity including anomaly
compensating terms.

   The realization of the similar program in the usual
supergravity requires the expression of the $T_{abc}$-field in
terms of $H_{abc}$.  For this purpose one must solve  the highly
nonlinear differential equation (3.9). It is possible to get
only  perturbative solution in the form of infinite series in
powers of $\gamma $-parameter. It is rather difficult to
calculate even the first term of this series (cf.
\cite{RRZ}).

In the framework of the discussed approach one can not calculate
all possible superstring corrections. Even the corrections
needed for compensation of anomalies were not taken into account
in the full scale. Indeed, the $N$-Bianchi Identities (3.1) were
not modified in accordance with the general procedure
\cite{GN3}, \cite{SS}.  It means that only tree-level
syperstring corrections are taken into account, namely the
corrections where  scale invariances \cite{GN3},
\cite{W}, specific for the tree-level, are not broken.  In
addition, one-loop anomaly compensating counter-terms, which are
not related with the structure of BI's, are out of the scope if
they break the scale invariance. The same approximation is
typical for the most papers on this subject. (Note, that there are some
reasons \cite{T2}
to consider  string-loop effects mentioned above as very small
numerically).

 \bigskip

 The author is thankful to S.Gates for information on the
H.Nishino paper \cite{N}.

\end{document}